\documentclass[preprint,nofootinbib]{revtex4}%
\usepackage{amssymb}
\usepackage{amsfonts}
\usepackage{amsmath}
\usepackage{graphicx}%
\begin{document}
\title{Microscopic entropy of the three-dimensional rotating black hole of BHT
massive gravity}
\author{Gaston Giribet$^{1}$, Julio Oliva$^{2,3}$, David Tempo$^{2,4,5}$ and Ricardo
Troncoso$^{2,6}$}
\affiliation{$^{1}$Center for Cosmology and Particle Physics, New York University, 4
Washington Place NY10003, New York, USA.}
\affiliation{$^{2}$Centro de Estudios Cient\'{\i}ficos (CECS), Casilla 1469, Valdivia,}
\affiliation{$^{3}$Instituto de F\'{\i}sica, Facultad de Ciencias, Universidad Austral de Chile.}
\affiliation{$^{4}$Departamento de F\'{\i}sica, Universidad de Concepci\'{o}n, Casilla,
160-C, Concepci\'{o}n, Chile.}
\affiliation{$^{5}$Physique th\'{e}orique et math\'{e}matique, Universit\'{e} Libre de
Bruxelles, ULB Campus Plaine C.P.231, B-1050 Bruxelles, Belgium, }
\affiliation{$^{6}$Centro de Ingenier\'{\i}a de la Innovaci\'{o}n del CECS (CIN), Valdivia, Chile.}
\preprint{CECS-PHY-09/07}

\begin{abstract}
Asymptotically AdS rotating black holes for the Bergshoeff-Hohm-Townsend (BHT)
massive gravity theory in three dimensions are considered. In the special case
when the theory admits a unique maximally symmetric solution, apart from the
mass and the angular momentum, the black hole is described by an independent
\textquotedblleft gravitational hair\textquotedblright\ parameter, which
provides a negative lower bound for the mass. This bound is saturated at the
extremal case and, since the temperature and the semiclassical entropy vanish,
it is naturally regarded as the ground state. The absence of a global charge
associated with the gravitational hair parameter reflects through the first
law of thermodynamics in the fact that the variation of this parameter can be
consistently reabsorbed by a shift of the global charges, giving further
support to consider the extremal case as the ground state. The rotating black
hole fits within relaxed asymptotic conditions as compared with the ones of
Brown and Henneaux, such that they are invariant under the standard asymptotic
symmetries spanned by two copies of the Virasoro generators, and the algebra
of the conserved charges acquires a central extension. Then it is shown that
Strominger's holographic computation for general relativity can also be
extended to the BHT theory; i.e., assuming that the quantum theory could be
consistently described by a dual conformal field theory at the boundary, the
black hole entropy can be microscopically computed from the asymptotic growth
of the number of states according to Cardy's formula, in exact agreement with
the semiclassical result.

\end{abstract}
\maketitle

\section{Introduction}

The new theory of massive gravity in three dimensions, recently proposed by
Bergshoeff, Hohm and Townsend (BHT) \cite{BHT}, has naturally earned a great
deal of attention since it enjoys many remarkable properties. The theory is
described by the parity-invariant action%
\begin{equation}
I_{BHT}=\frac{1}{16\pi G}\int d^{3}x\sqrt{-g}\left[  R-2\lambda-\frac{1}%
{m^{2}}\left(  R_{\mu\nu}R^{\mu\nu}-\frac{3}{8}R^{2}\right)  \right]  \ ,
\label{Action}%
\end{equation}
which yields fourth order field equations for the metric. Noteworthy, since at
the linearized level they are equivalent to the Fierz-Pauli equations for a
massive spin-2 field, ghosts are \textquotedblleft exorcized'' from the theory
\cite{BHT,Deser-Alas, BHT2}. As a consequence, the BHT theory appears to be
unitary \cite{Unitarity} and renormalizable \cite{Renormalizable}. A variety
of exact solutions has been found \cite{Clement 1, ABGH, BHT2, OTT, Kundt
spacetimes, ABGGH}, its locally supersymmetric extension is known
\cite{Supergravity}, and further aspects have been developed in \cite{further
aspects}.

In the special case, $m^{2}=\lambda$, the theory possesses a unique maximally
symmetric solution and it acquires additional interesting features, as it is
the enhancement of gauge invariance for the linearized theory, such that the
graviton is described by a single degree of freedom \cite{BHT2} being
\textquotedblleft partially massless\textquotedblright%
\ \cite{Deser-Nepomechie, Deser-Waldron-1, Deser-Waldron-2, Tekin}. For the
nonlinear theory this is reflected in the fact that the AdS waves propagate a
single scalar degree of freedom whose mass saturates the
Breitenlohner-Freedman bound \cite{ABGH}. It is also known that in this case,
the Brown-Henneaux boundary conditions can be consistently relaxed, which
enlarges the space of admissible solutions so as to include rotating black
holes, gravitational solitons, kinks and wormholes \cite{OTT}.

In what follows we will focus on the asymptotically AdS rotating black hole
found in \cite{OTT}. The solution is described in terms of two global charges,
being the mass and the angular momentum, as well as by an additional
"gravitational hair" parameter, which provides a negative lower bound for the
mass. This bound is saturated at the extremal case and, since the temperature
and the semiclassical entropy vanish, it is naturally regarded as the ground
state. As revisited in the next Section, this sort of extremality is due to
the gravitational hair and it turns out to be stronger than extremality due to
rotation. In Section \ref{1st law} it is shown that the absence of a global
charge associated with the gravitational hair parameter is reflected in the
first law of thermodynamics through the fact that the variation of this
parameter can be consistently reabsorbed by a shift of the global charges,
giving a remarkably strong support to consider the extremal case as the ground
state. Since the rotating black hole fits within relaxed asymptotic conditions
as compared with the ones of Brown and Henneaux \cite{Brown-Henneaux}, such
that they are invariant under the standard asymptotic symmetries spanned by
two copies of the Virasoro generators, and the algebra of the conserved
charges acquires a central extension, Section \ref{Strominger+Cardy} is
devoted to show that Strominger's holographic result for general relativity
\cite{Strominger-Cardy} can also be extended to the BHT theory; i.e., assuming
that the quantum theory could be consistently described by a dual conformal
field theory at the boundary, the black hole entropy can be microscopically
computed from the asymptotic growth of the number of states according to
Cardy's formula, in exact agreement with the semiclassical result. Ending
remarks are made in Section \ref{Discussion and comments}.

\section{Rotating black hole}

\label{Section Rotating black hole}

The BHT\ theory (\ref{Action}) for the special case, $m^{2}=\lambda=-\frac
{1}{2l^{2}}$, admits the following rotating black hole solution \cite{OTT}%

\begin{equation}
ds^{2}=-NFdt^{2}+\frac{dr^{2}}{F}+r^{2}\left(  d\phi+N^{\phi}dt\right)
^{2}\ , \label{Rotating black hole metric}%
\end{equation}
where $N$, $N^{\phi}$ and $F$ are functions of the radial coordinate $r$,
given by%
\begin{align}
N  &  =\left[  1+\frac{bl^{2}}{4H}\left(  1-\Xi^{\frac{1}{2}}\right)  \right]
^{2}\ ,\nonumber\\
N^{\phi}  &  =-\frac{a}{2r^{2}}\left(  4GM-bH\right)  \ ,\label{Ns&F}\\
F  &  =\frac{H^{2}}{r^{2}}\left[  \frac{H^{2}}{l^{2}}+\frac{b}{2}\left(
1+\Xi^{\frac{1}{2}}\right)  H+\frac{b^{2}l^{2}}{16}\left(  1-\Xi^{\frac{1}{2}%
}\right)  ^{2}-4GM\ \Xi^{\frac{1}{2}}\right]  \ ,\nonumber
\end{align}
and
\begin{equation}
H=\left[  r^{2}-2GMl^{2}\left(  1-\Xi^{\frac{1}{2}}\right)  -\frac{b^{2}l^{4}%
}{16}\left(  1-\Xi^{\frac{1}{2}}\right)  ^{2}\right]  ^{\frac{1}{2}}\ .
\label{H}%
\end{equation}
Here $\Xi:=1-a^{2}/l^{2}$, and the rotation parameter $a$ is bounded in terms
of the AdS radius according to $-l\leq a\leq l$. The solution is then
described by two global charges, where $M$ is the mass and $J=Ma$ is the
angular momentum, as well as by an additional\footnote{For simplicity, here
the gravitational hair parameter $b$ has been redefined making $b\rightarrow
b\Xi^{1/2}$ in \cite{OTT}.} "gravitational hair" parameter $b$.

The rotating black hole is a conformally flat asymptotically AdS spacetime,
and depending on the range of the parameters $M$, $a$ and $b$, the solution
possesses an ergosphere and a singularity that can be surrounded by event and
inner horizons. In the case of $b=0$, the solution reduces to the BTZ black
hole \cite{BTZ, BHTZ}, while when the gravitational hair parameter is switched
on ($b\neq0)$, the spacetime is no longer of constant curvature and the
solutions splits in two branches according to the sign of $b$. The event
horizon radius, the temperature and the entropy are given by $r_{+}=\gamma
\bar{r}_{+}$, $T=\gamma^{-1}\bar{T}$, and $S=\gamma\bar{S}$, respectively,
where $\gamma^{2}=\frac{1}{2}\left(  1+\Xi^{-1/2}\right)  $, and $\bar{r}_{+}%
$,$\bar{T}$, $\bar{S}$ correspond to the radius of the event horizon, the
temperature and the entropy for the static case. Thus, the angular velocity of
the horizon turns out to be%
\begin{equation}
\Omega_{+}=\frac{1}{a}\left(  \Xi^{\frac{1}{2}}-1\right)  \ , \label{Omega+}%
\end{equation}
and the Hawking temperature and the Entropy can be explicitly expressed as%
\begin{align}
T  &  =\frac{1}{\pi l}\Xi^{\frac{1}{2}}\sqrt{2G\Delta M\left(  1+\Xi^{\frac
{1}{2}}\right)  ^{-1}}\ ,\label{Temperature}\\
S  &  =\pi l\sqrt{\frac{2}{G}\Delta M\left(  1+\Xi^{\frac{1}{2}}\right)  }\ ,
\label{Entropy}%
\end{align}
where
\begin{equation}
\Delta M:=M-M_{0}=M+\frac{b^{2}l^{2}}{16G}\ .
\end{equation}
Note that the rotating BTZ black hole ($b=0$) possesses twice the entropy
obtained from general relativity, i.e., $S=\frac{A_{+}}{2G}$.

The black hole described by (\ref{Rotating black hole metric}) fulfills%
\begin{equation}
M^{2}\geq\frac{J^{2}}{l^{2}}\ . \label{Rotation bound}%
\end{equation}
This bound is saturated when the rotation parameter is given by $a^{2}=l^{2}$,
so that the angular velocity of the horizon is $\Omega_{+}^{2}=\frac{1}{l^{2}%
}$ and the temperature (\ref{Temperature}) vanishes. This is an extremal case
since the event and inner horizons coincide, and for $b\neq0$ they are on top
of the singularity which become null and it is located at%
\begin{equation}
r_{+}^{2}=r_{-}^{2}=r_{s}^{2}=2Gl^{2}\Delta M\ .
\end{equation}
Note that for $a^{2}=l^{2}$ the entropy (\ref{Entropy}) reduces to $S=\pi
l\sqrt{\frac{2}{G}\Delta M}$.

The case $b<0$ is particularly interesting since the black hole mass is
allowed to be negative up to certain extent, and it is bounded in terms of the
gravitational hair parameter according to
\begin{equation}
M\geq M_{0}\ ,\label{Bound}%
\end{equation}
with
\begin{equation}
M_{0}=-\frac{b^{2}l^{2}}{16G}\ .\label{M0}%
\end{equation}
This opens the possibility of having a different kind of stronger extremality.
Indeed, the bound (\ref{Bound}) is saturated in the case of $M=M_{0}$, so that
the metric describes an extremal black hole for which the event and the inner
horizons coincide
\[
r_{+}^{2}=r_{-}^{2}=\frac{b^{2}l^{4}}{8}\Xi^{\frac{1}{2}}\left(  1+\Xi
^{\frac{1}{2}}\right)  \ ,
\]
always enclosing a timelike singularity located at
\begin{equation}
r_{s}^{2}=\frac{b^{2}l^{4}}{8}\Xi^{\frac{1}{2}}\left(  \Xi^{\frac{1}{2}%
}-1\right)  .
\end{equation}
Remarkably, for $M=M_{0}$, not only the temperature but also the entropy
vanishes, as it is shown by Eqs. (\ref{Temperature}) and (\ref{Entropy}).
Thus, it is natural to regard the case of $M=M_{0}$ as the ground state, not
only because it is the lower bound for the mass allowed by cosmic censorship,
but also because, since the entropy vanishes, it would correspond to a single
nondegenerate microscopic state.

Note than this kind extremality is due to the existence of the gravitational
hair parameter and it can be attained for any value of the rotation parameter
$a$ within its allowed range, so that the angular momentum is $J_{0}=M_{0}a$,
and the extremal horizon has an angular velocity given by (\ref{Omega+}).

As explained in Section \ref{Strominger+Cardy}, at the extremal case $M=M_{0}%
$, not only the entropy, but also both left and right temperatures vanish,
while for the extremal case $J^{2}=M^{2}l^{2}$, only one of this temperatures
vanishes, let say the left, while neither the right temperature nor the
entropy do. Thus, also this sense, extremality due to gravitational hair is
stronger than extremality due to rotation.

\section{Gravitational hair, first law of thermodynamics and the ground state}

\label{1st law}

Following the Deser-Tekin approach \cite{Deser-Tekin}, it was shown in
\cite{OTT} that the rotating black hole (\ref{Rotating black hole metric})
possesses only two global charges, the mass $M$ and the angular momentum
$J=Ma$, where the massless BTZ black hole was chosen as the reference
background. Thus, because of the absence of a global charge associated to $b$,
it was dubbed as the gravitational hair parameter.

The absence of a global charge associated with $b$ is then reflected in the
first law of thermodynamics through the fact that no chemical potential can be
associated to it, and hence the variation of this parameter has to be
consistently reabsorbed by a shift of the global charges.

This can be explicitly seen as follows: According to Eqs. (\ref{Temperature})
and (\ref{Entropy}), the product of the temperature and the variation of the
entropy is given by%

\[
TdS=\Xi^{\frac{1}{2}}dM+\frac{bl^{2}}{8G}\Xi^{\frac{1}{2}}db-\frac{1}%
{a}\left(  1-\Xi^{\frac{1}{2}}\right)  \left(  M+\frac{b^{2}l^{2}}%
{16G}\right)  da\ ,
\]
and taking into account Eqs. (\ref{Omega+}) and (\ref{M0}), this equation
reduces to%
\begin{equation}
d\left(  M-M_{0}\right)  =TdS-\Omega_{+}d\left(  J-J_{0}\right)  \ ,
\label{First law}%
\end{equation}
where $M_{0}$ and $J_{0}=M_{0}a$ correspond to the mass and the angular
momentum of the extremal case, respectively.

\bigskip

As expected, the dependence on the gravitational hair parameter is entirely
absorbed by a shift of the global charges. Remarkably, Eq. (\ref{First law})
means that the shift is precisely such that the first law is fulfilled
provided the global charges (the mass and the angular momentum) are measured
with respect to the ones of the extremal case that saturates the bound
(\ref{Bound}).

This provides further strong support to consider the extremal case as the
ground state. Using this fact, in the next section it is shown that the
entropy of the rotating black hole (\ref{Entropy}) can be microscopically
computed from the asymptotic growth of the number of states of the dual theory.

\section{Microscopic entropy of the rotating black hole}

\label{Strominger+Cardy}

As shown in \cite{OTT}, the rotating black hole
(\ref{Rotating black hole metric}) fits within a relaxed set of asymptotic
conditions as compared with the one of Brown and Henneaux
\cite{Brown-Henneaux}, being such that they are invariant under the standard
asymptotic symmetries spanned by two copies of the Virasoro generators. The
algebra of the conserved charges also acquires a central extension being twice
the value found for general relativity, i.e.,%
\begin{equation}
c_{\pm}=c=\frac{3l}{G}\ .\label{central charge}%
\end{equation}
Choosing the extremal case as the reference background, the only nonvanishing
surface integrals for the rotating black hole are then the ones associated
with the left and right Virasoro generators $L_{0}^{\pm}$, given by%
\begin{equation}
\Delta_{\pm}=\frac{1}{2}\left(  l\Delta M\pm\Delta J\right)  =\frac{1}%
{2}\Delta M\left(  l\pm a\right)  \ .\label{Delta+-}%
\end{equation}
where $\Delta M=M-M_{0}$, and $\Delta J=\Delta Ma$, are mass and the angular momentum.

Regarding this as the starting point, one can see that Strominger's result for
general relativity \cite{Strominger-Cardy} works also for the BHT theory
described by (\ref{Action}). Strominger holographic computation relies on an
observation pushed forward more than twenty years ago by Brown and Henneaux
\cite{Brown-Henneaux}, who suggested that since asymptotic symmetry group of
general relativity with negative cosmological constant in three dimensions is
generated by two copies of the Virasoro algebra, a consistent quantum theory
of gravity should be described terms of a two-dimensional conformal field
theory. This is currently interpreted in terms of the AdS/CFT correspondence
\cite{Maldacena-Klebanov-Witten}.

Hence, assuming that quantum theory for BHT\ massive gravity exists and it is
consistently described by a dual conformal field theory at the boundary, the
physical states must form a representation of the algebra with a central
charge given by (\ref{central charge}), and if the CFT fulfills some
physically sensible properties, the asymptotic growth of the number of states
is given by Cardy's formula.

Therefore, as explained above, since the black hole
(\ref{Rotating black hole metric}) can be regarded as excitations of the
ground state, which corresponds to the extremal case $M=M_{0}$, the entropy
can be computed in the microcanonical ensemble as the logarithm of the density
of states, given by
\begin{equation}
S=2\pi\sqrt{\frac{c_{+}}{6}\Delta_{+}}+2\pi\sqrt{\frac{c_{-}}{6}\Delta_{-}%
}\ ,\label{Cardy formula}%
\end{equation}
where $c_{\pm}$ is given by (\ref{central charge}), and $\Delta_{\pm}$ in Eq.
(\ref{Delta+-}) correspond to the eigenvalues of $L_{0}^{\pm}$. Thus, Eq.
(\ref{Cardy formula}) reduces to
\begin{align}
S &  =\pi l\sqrt{\frac{\Delta M}{G}}\left(  \sqrt{1+\frac{a}{l}}+\sqrt
{1-\frac{a}{l}}\right)  \ ,\\
&  =\pi l\sqrt{\frac{2}{G}\left(  1+\Xi^{\frac{1}{2}}\right)  \Delta M}\ ,
\end{align}
which exactly agrees with the semiclassical result in Eq. (\ref{Entropy}).

\bigskip

Note that since left and right movers are decoupled, they can be at
equilibrium at different temperatures $T_{\pm}$. In the canonical ensemble,
the entropy acquires the form%
\begin{equation}
S=\frac{\pi^{2}l}{3}\left(  c_{+}T_{+}+c_{-}T_{-}\right)  \ ,
\label{Cardy with T+-}%
\end{equation}
and since the free energy is given by
\begin{equation}
F=\left(  \beta_{+}\Delta_{+}+\beta_{-}\Delta_{-}\right)  l^{-1}-S=\beta\Delta
M+\beta\Omega_{+}\Delta J-S\ ,
\end{equation}
the left and right temperatures turn out to be%
\begin{equation}
T_{\pm}=\frac{T}{1\pm l\Omega_{+}}=\frac{1}{2\pi l}\left(  1\mp l\Omega
_{+}\right)  \sqrt{2G\Delta M\left(  1+\Xi^{\frac{1}{2}}\right)  }\ .
\label{T+-}%
\end{equation}
Then, by virtue of Eqs. (\ref{central charge}) and (\ref{T+-}) it is simple to
verify that formula (\ref{Cardy with T+-}) exactly reproduces the black hole
entropy (\ref{Entropy}) as well.

Note that for extremal black holes case due to rotation, $J^{2}=Ml^{2}$, for
which $\Omega_{+}^{2}=\frac{1}{l^{2}}$, only one of this temperatures
vanishes, let say the left, while the right temperature is given by
$T_{+}=\frac{1}{\pi l}\sqrt{2G\Delta M\left(  1+\Xi^{\frac{1}{2}}\right)  }$
and they have a nonvanishing entropy $S=\pi l\sqrt{\frac{2}{G}\Delta M}$. It
is reassuring then to verify that for the extremal case due to gravitational
hair, $M=M_{0}$, not only the entropy, but also both left and right
temperatures vanish, as it has to be for a suitable ground state.

\section{Discussion and comments}

\label{Discussion and comments}

It was shown that the semiclassical entropy of the rotating black hole
(\ref{Rotating black hole metric}) can be microscopically reproduced from
Cardy's formula (\ref{Cardy formula}), where the ground state turns out to be
given by the extremal case $M=M_{0}$. It is worth pointing out that the
computations can be extended perfectly well even for a case that they were not
intended for, $b>0$. The subtlety is related to the fact that for $b>0$, the
black hole configuration with $M=M_{0}$ suffers certain pathologies.
Nevertheless, as it was shown in \cite{OTT}, in this case the theory also
admits a gravitational soliton for $M=M_{0}$. Thus, since the spacetime is
regular everywhere, the soliton provide a suitable nondegenerate state that
can be naturally regarded as the ground state. This point is left for future
detailed discussion.

Since the rotating black hole (\ref{Rotating black hole metric}) is
conformally flat, it solves the BHT field equations for the special case,
$m^{2}=\lambda$, even in presence of the topological mass term, and it is
simple to verify that our results extend to this case. The vanishing of the
Cotton tensor should also imply that the rotating black hole is conformally
related to the matching of different solutions of constant curvature by means
of an improper conformal transformation, as it occurs for the static case
\cite{Joao Pessoa 2+1}.

\bigskip

\textit{Acknowledgments.} We thank D. Anninos, and S. Detournay and M.
Henneaux for useful discussions and comments. G.G. is Member of Conicet,
Argentina, on leave of absence from University of Buenos Aires. D.T. thanks
Conicyt for financial support. This research is partially funded by Fondecyt
grants N%
${{}^o}$
1061291, 1071125, 1085322, 1095098, 3085043, and by ANPCyT grant
PICT-2007-00849. The Centro de Estudios Cient\'{\i}ficos (CECS) is funded by
the Chilean Government through the Millennium Science Initiative and the
Centers of Excellence Base Financing Program of Conicyt. CECS is also
supported by a group of private companies which at present includes
Antofagasta Minerals, Arauco, Empresas CMPC, Indura, Naviera Ultragas and
Telef\'{o}nica del Sur. CIN is funded by Conicyt and the Gobierno Regional de
Los R\'{\i}os.

\end{document}